\documentclass[aps,prd,preprint,preprintnumbers,nofootinbib,floatfix]{revtex4}
\usepackage{epsfig}
\usepackage{latexsym,amsmath,amssymb,amstext}
\usepackage{float}
\usepackage{graphicx,xcolor}
\usepackage[paperwidth=210mm,paperheight=297mm,centering,hmargin=2cm,vmargin=2.5cm]{geometry}

\newcommand{\be}{\begin{equation}}
\newcommand{\ee}{\end{equation}}
\newcommand{\bea}{\begin{eqnarray}}
\newcommand{\eea}{\end{eqnarray}}
\newcommand{\Tr}{{\rm Tr}}

\newcommand{\pathi}{{\cal P}\!\!\!\!\!\!\int}

\newcommand\bef{\begin{figure}}
\newcommand\eef[1]{\label{fg:#1}\end{figure}}
\newcommand\beq{\begin{equation}}
\newcommand\eeq[1]{\label{#1}\end{equation}}
\newcommand\beqa{\begin{eqnarray}}
\newcommand\eeqa[1]{\label{#1}\end{eqnarray}}
\newcommand\bet{\begin{table}}
\newcommand\eet[1]{\label{tb:#1}\end{table}}

\begin{document}
\title{Topological gauge actions on the lattice as Overlap fermion determinants}
\preprint{JLAB-THY-22-3591}
\author{Nikhil\ \surname{Karthik}}
\email{nkarthik.work@gmail.com}
\affiliation{Department of Physics, College of William \& Mary, Williamsburg, VA 23185, USA}
\affiliation{Thomas Jefferson National Accelerator Facility, Newport News, VA 23606, USA}
\author{Rajamani\ \surname{Narayanan}}
\email{rajamani.narayanan@fiu.edu}
\affiliation{Department of Physics, Florida International University, Miami, FL 33199}

\begin{abstract}

Overlap fermion on the lattice has been shown to properly reproduce topological aspects of gauge fields. In this paper, we 
review the derivation of Overlap fermion formalism in 
a torus of three space-time dimensions. Using the formalism, we 
show how to use the 
Overlap fermion determinants in the massless and infinite mass limits to construct 
different continuum topological gauge actions, 
such as the level-$k$ Chern-Simons action, ``half-CS" term and the 
mixed Chern-Simons (BF) coupling,
in a 
gauge-invariant lattice UV regulated manner. Taking 
special Abelian and non-Abelian background fields, 
we demonstrate numerically how the lattice formalism 
beautifully reproduces the continuum expectations, 
such as the flow of action under large gauge transformations.

\end{abstract}
\maketitle

\section{Introduction}

The gauge theories in three space-time dimensions admit a parity-odd 
Chern--Simons (CS) topological gauge action in addition to the parity-even 
Maxwell gauge action. The Maxwell 
theory can be nonperturbatively regulated via the lattice discretization of space-time and by using the 
local plaquette gauge action. The CS theories
are not so straightforward to regulate on the lattice, 
mainly due to the fact that the CS action
is only gauge-invariant up to integer winding under nontrivial gauge 
transformations (e.g., \cite{Dunne:1998qy}) and it is not possible to realize such a term simply as a 
local Wilson loop gauge action. Vigorous research work is being 
conducted on CS theories coupled to matter content and certain
infrared duality relations \cite{Seiberg:2016gmd,Karch:2016sxi,Wang:2017txt} have been conjectured to exist
at critical points separating different topological phases. 
Therefore, the question of how to study such theories numerically 
on the lattice is important.
The aim of this paper is to elucidate how to introduce topological 
gauge actions, such as the Chern--Simons action, on the lattice in a
completely gauge-invariant manner by identifying such actions as the 
induced gauge actions of \mbox{lattice fermions.}

Let us first consider gauge theories in even dimensions to see how gauge field topology is realized using lattice fermions. The space of Euclidean continuum gauge fields, \mbox{$A = A_\mu(x) d x_\mu$,} in even dimensional space, $D=2m$, usually has infinitely many disconnected pieces and each piece has an associated topological number~\footnote{ This is well known and a chapter or more is attributed to this topic in all modern books on quantum field theory; we find it useful to refer to the lecture notes by Bilal~\cite{Bilal:2008qx} which has a complete self-contained description and has citations to other relevant lecture notes and books.} 
given by
\be
Q = \frac{1}{m! (2\pi)^m} \int P_m(F);\qquad P_m(F)=\Tr F^m;
\ee
where 
$F = dA + iA\wedge A$ 
is the Euclidean field strength associated with $A_\mu(x)$ and $F^m = F\wedge F\cdots \wedge F$.
 As such not all gauge fields can be connected to the trivial one, $A_\mu(x)=0$. 
 One way to nonperturbatively regularize a gauge theory is 
 using lattice, where one introduces gauge fields via gauge-links
 that connect neighboring lattice sites.
 Link variables belonging to the Lie group defined by the path ordered product of the Lie group elements,
\be
U_\mu(x) = \pathi_x^{x+\hat\mu} e^{iA_\mu(y)} dy 
\equiv e^{i A^{\rm lat}_\mu(x)}
\ee
 along the path connecting $x$ and $x+\hat\mu$ (we have set the lattice spacing to unity and $x$ takes on integer values) are lattice gauge fields. Naively, $U^q_\mu(x) = e^{iqA^{\rm lat}_\mu(x)}$ for some real valued 
 parameter $q$, continuously connects any gauge field configuration on the lattice to the trivial one, $U_\mu(x)=1$, by sliding the value of 
 $q$ from 0 to 1 {\sl seemingly}
 without encountering any singular behavior in gauge-links or 
 the plaquettes at any $x$ 
 during the process.
 Notwithstanding the apparent lack of discontinuity on the lattice 
 between any
 two gauge-fields that could otherwise be topologically distinct 
 from each other in the 
 continuum,
an assignment of a topological integer to every gauge field configuration is still possible. A straightforward approach is to invoke the Atiyah--Singer index theorem \cite{Atiyah:1963zz} and use fermions to match $Q$ with the index of a lattice Dirac operator. For every lattice gauge field background in even dimensions and the associated massive Hermitian Wilson--Dirac operator, $H_w(U;m_w)$, the index is the difference between the total number of negative eigenvalues of $H_w(U;\pm m_w)$ \cite{Narayanan:1994gw}. If the index associated with a particular, $U_\mu(x)=e^{iA_\mu(x)}$ is not zero, we will see an eigenvalue of $H_w(U^q; m_w)$ cross zero as one smoothly changes $q\in [0,1]$ in $U^q_\mu(x) = e^{iqA_\mu(x)}$. Therefore, there is one value of $q$ where the ground state of the many body operator 
 \be
 {\cal H}_w(U;m_w) = a^\dagger H_w(U;m_w) a,
 \ee
 for a $D+1$ dimensional auxiliary fermionic system,
 with $a^\dagger$, $a$ being canonical fermion creation and annihilation operators, is doubly degenerate. As is also well known, chiral gauge anomalies in even dimensions are closely related to the topological index \cite{Bilal:2008qx} and this 
 can also be understood in terms of
 the ground state, $|0;U;m_w\rangle$, of ${\cal H}_w(U;m_w)$
 as explained in \cite{Neuberger:1998xn} .
 Having defined the one form,
 \be
 d|0;U;m_w\rangle = \sum_{\mu,x} \frac{\partial}{\partial A_\mu(x)} |0;U;m_w\rangle d A_\mu(x)
 \ee
 it is shown in \cite{Neuberger:1998xn} that
 \be
 d\left[ j^{\rm cons}-j^{\rm cov}\right] = \Tr \left [ P dP \wedge dP\right];\qquad P = |0;U;m_w\rangle\langle 0;U;m_w| \label{deltaj}
 \ee
 is a well defined function of the lattice gauge field background and $j^{\rm cons}$ and $j^{\rm cov}$ are the consistent and covariant currents.
The problem of anomaly cancellation can be studied using Equation \eqref{deltaj} and the need to fine tune the lattice Wilson--Dirac operator is discussed in \cite{Neuberger:1998xn}.
The above discussion on the ability of 
massless overlap fermion to detect and classify topologically distinct 
gauge sectors on the lattice is well-known. In this paper, we 
review the aspects of overlap fermions in odd-dimensions, especially in 
2 + 1 dimensions, and how the parity anomaly of overlap fermions can be used to introduce topological gauge 
actions that are characteristic of odd-dimensional gauge theories.

Chiral anomaly inducing topological index in even dimensions and parity anomaly inducing the Chern--Simons action in odd-dimensions are locally related as \cite{Bilal:2008qx}
\be
P_m(F) = dQ_{2m-1}(A,F)
\ee
where $Q_{2m-1}(A,F)$ is the Chern--Simons form in one dimension lower, namely, $D=2m-1$. Setting a one-parameter family of gauge fields equal to $A_t =t A$,
and noting that $F_t=tdA + it^2 A\wedge A$, 
\be
Q_{2m-1}(A,F) = m \int_0^1 dt \ \Tr (A \wedge F_t^{m-1});\qquad m > 1.
\ee

Focusing on $m=2$, we have
\be
Q_3(A,F) = \Tr \left[A\wedge dA + i\frac{2}{3} A\wedge A \wedge A\right].
\ee

Similar to our discussion on the challenge in defining
the topological index simply as a local operator constructed out of
local Wilson-loop operators on the lattice, it is not simple to 
define the above Chern--Simons form as a local gauge-link-based operator
and be able to satisfy invariance under large gauge transformations of 
the type we will discuss later in this paper. Solution to this problem
again is to introduce the Chern--Simons action using the fermions on 
the lattice; concretely, through the parity-odd part of the 
induced gauge action from overlap fermions.
An early study in Ref. \cite{Coste:1989wf} showed that 
the Abelian parity anomaly is reproduced using lattice perturbation theory with a single-flavor of two-component Wilson fermion with non-zero mass at lattice UV scales \cite{Coste:1989wf}.
The important point we stress in this paper is that the 
massive two-component Wilson Dirac operator $X$ on any 
background field $A^{\rm latt}$, immediately 
leads to a gauge covariant unitary operator \cite{Kikukawa:1997qh}, $V$,
\be
V \equiv \left(X X^\dagger\right)^{-1/2} X,
\ee
and the gauge-invariant phase of $\det(V)$ is parity-odd and becomes 
the lattice realization of the Chern--Simons action for any gauge field
background \cite{Kikukawa:1997qh,Narayanan:1997by,Bietenholz:2000ca,Bietenholz:2002mt}. The unitary operator $V$ is nothing but the overlap
operator of a two-component fermion of mass of inverse lattice spacing.
The phase within lattice regularization has been extensively analyzed in \cite{Karthik:2015sza} for various Abelian backgrounds. 
In addition to the Chern--Simons action, the recent literature on fractional quantum Hall states rely heavily on parity-anomalous 
two-component exactly massless Dirac fermions that leads to the 
so-called ``half-Chern--Simons" term. Subtleties arise when discussing half the Chern--Simons action while maintaining gauge invariance \cite{AlvarezGaume:1984nf,Witten:2015aoa,Seiberg:2016gmd}. 
We also show how the construction of the unitary lattice operator $V$ 
also immediately leads to the generalization of the Chern--Simons 
term to include the BF terms such as $B\wedge F = B\wedge dA$.

In order to keep this paper as self-contained as possible, we first 
review the derivation and the salient features of overlap fermions in three dimensions in Section \ref{sec2}. In Section \ref{sec3}, we focus on the  variation of overlap fermion determinant as fermion mass is varied 
from $1/a$ to massless limit; the point of this discussion is 
to show that the infinite mass and zero fermion mass limits indeed 
correctly reproduce the Chern--Simons and ``half-Chern--Simons'' terms correctly in the continuum limit and independent of any lattice UV regulator parameters, such as the mass term in the 
Wilson fermion kernel. More interestingly, in Section \ref{sec4}, we take specific Abelian backgrounds with non-trivial topology on 2d spatial planes and show 
how the flow from infinite mass to zero mass limit preserves 
gauge invariance. For this, we follow the discussion in \cite{Kikukawa:1997qh}. In Section \ref{sec5}, we take a non-Abelian background to
discuss how the $A\wedge A \wedge A$ part of CS term present for 
non-Abelian case is correctly reproduced. After the discussion of
the Chern--Simons terms, in Section \ref{sec6}, we focus on straight-forward extensions of 
overlap formalism to implement mixed Chern--Simons terms that couple 
two different gauge field backgrounds, and as a consequence, 
provide dictionary between some of the recently proposed 
fermion-boson dualities in the continuum to those on the 
lattice.

\section{Overlap Formalism in Three Dimensions} \label{sec2}

This section follows \cite{Karthik:2016ppr} very closely and we repeat the derivation while keeping a phase ambiguity intact till the very end. 
Despite this paper being about nonperturbative 
regularization of topological field theories, the 
lattice formalism is strictly presented on toroidal $S\times S \times S$
manifold tessellated into uniform cubes of volume $a^3$, 
with $a$ being the lattice spacing.
The na\"ive massless Dirac operator on a three dimensional lattice (we will set the lattice spacing to unity) is given by
\be
D = \frac{1}{2} \sum_{\mu=1}^3 \sigma_\mu \left( T_\mu - T_\mu^\dagger\right);\ \ \ 
D^\dagger = -D;\label{naive}
\ee
where $\sigma_\mu$ are Pauli matrices satisfying $\sigma_\mu \sigma_\nu = \delta_{\mu\nu}+ i \epsilon_{\mu\nu\lambda}\sigma_\lambda$, 
and the 
action of translation operator $(T_\mu \psi)(x) = U_\mu(x) \psi(x+\hat\mu)$ up to
lattice periodicity.
Under parity ($x_\mu \to - x_\mu$,
\be
T_\mu \to T_\mu^\dagger\quad \rightarrow \quad D\to -D.
\ee
and under a gauge transformation ${\cal G}$, 
\be
T_\mu \to {\cal G} T_\mu {\cal G}^\dagger;\ \ \ \ \
({\cal G}\phi)(x) = g(x)\phi(x);\ \ \ \ {\cal GG}^\dagger =1,
\ee
which implies
\be
D \to {\cal G} D {\cal G}^\dagger.
\ee

The na\"ive massless Dirac operator has a two fold degeneracy in all gauge field backgrounds. Furthermore, for every eigenvalue there is one with the opposite sign. To see these two features, we observe that the anti-Hermitian operator only couples odd lattice sites with even lattice sites. The eigenvalues come in $\pm i\lambda$ pairs and the fermion determinant is real and positive in all gauge backgrounds and there is no parity anomaly.
In order to realize a single flavor two-component massive Dirac fermion without any doublers
in the overlap formalism \cite{Narayanan:1994gw}, we define two Hamiltonians
that act on four component spinors:
\be
H_-=\begin{pmatrix} 1 & 0 \cr 0 & -1\end{pmatrix};\qquad H_+ = \begin{pmatrix} B & D \cr -D & -B \end{pmatrix},
\ee
where $1$ denotes an identity matrix of the same size as $D$.
We have added the Wilson term,
\be
B =\frac{1}{2} \sum_{\mu=1}^3 \left( 2 - T_\mu - T_\mu^\dagger\right)-m_w;\ \ \ \
B=B^\dagger,\label{wilson}
\ee
with a Wilson mass parameter $0 < m_w < 2$ and $B\to B$ under parity. Under a gauge transformation
\be
B \to {\cal G}B{\cal G}^\dagger;\qquad H_+ \to {\cal G}H_+{\cal G}^\dagger.
\ee

Define the many body Hamiltonians by
\be
{\cal H}_{\pm} = -\begin{pmatrix} a^\dagger & b^\dagger\end{pmatrix} H_\pm \begin{pmatrix} a\cr b \end{pmatrix}
\ee
with $a^\dagger,b^\dagger$ and $a,b$ being canonical creation and annihilation operators for fermions.
With $|0\pm\rangle$ denoting the ground states of ${\cal H}_\pm$, 
the generating functional for a single two-component overlap fermion with a mass, $m$, is
\bea
Z(\eta,\bar\eta) &=& \langle 0-| \exp\left[ \bar\eta b + a^\dagger\eta + m a^\dagger b \right]|0+\rangle \cr
&=&
\int d\bar\xi d\xi e^{-\bar\xi \xi } \langle 0-| \exp\left[ \bar\chi b + a^\dagger\chi \right]|0+\rangle;\qquad\bar\chi = \bar\eta +m\bar\xi;\qquad \chi = \eta+\xi,
\eea
where $\eta,\bar\eta,\xi,\bar\xi$ are Grassmann variables.

The problem of diagonalizing $H_+$ in three dimensions is simplified by going to a new basis.
Let
\be
\Sigma = \frac{1}{\sqrt{2}} \begin{pmatrix} 1 & 1 \cr 1 & -1 \cr\end{pmatrix};\ \ \ 
\Sigma=\Sigma^\dagger;\ \ \ \ \Sigma^2=1.
\ee

The rotated Hamiltonian is
\be
H'_+=\Sigma H_+ \Sigma = \begin{pmatrix} 0 & B-D \cr B+D & 0\cr\end{pmatrix} \equiv \begin{pmatrix} 0 & X^\dagger \cr X & 0 \end{pmatrix}.
\ee

We can write
\be
X = L\Lambda R^\dagger;\qquad \Lambda_{ij} = \lambda_i \delta_{ij};\quad \lambda_i >0;\qquad R^\dagger R = L^\dagger L =1.
\ee

We define the unitary operator
as
\be
V = LR^\dagger = \frac{1}{\sqrt{XX^\dagger}} X\label{voper}
\ee
and does not suffer from the phase ambiguity present in $R$ ($L$ is fixed once $R$ is fixed).
Under parity,
\be
X\to X^\dagger; \qquad V \to V^\dagger
\ee
and under 
a gauge transformation
\be
X \to {\cal G} X {\cal G}^\dagger;\qquad V\to {\cal G} V {\cal G}^\dagger.
\ee

Let us make the dependence of $V$ on $U$ explicit and derive the relation under charge conjugation ($U \to U^*$):
\be
V(U^*) = \sigma_2 V^t(U) \sigma_2.\label{vccong}
\ee

We first note that 
\be
T^*_\mu(U) = T_\mu(U^*)\quad\Rightarrow\quad
B^*(U) = \sigma_2 B(U^*) \sigma_2;\qquad D^*(U) = -\sigma_2 D(U^*) \sigma_2.
\ee

From this we obtain
\be
X(U^*) = \sigma_2 X^t(U) \sigma_2,\quad\Rightarrow\quad X^\dagger(U^*) X(U^*) = \sigma_2 X^*(U) X^t(U) \sigma_2,
\ee
and our relation, Equation \eqref{vccong}, follows.

We can diagonalize $H_+$ as
\be
H_+ = {\cal U} \begin{pmatrix} \Lambda & 0 \cr 0 & - \Lambda\end{pmatrix} {\cal U}^\dagger; \qquad {\cal U} = \frac{1}{2} \begin{pmatrix} R+L & R-L \cr R-L & R+L
\end{pmatrix};\qquad {\cal UU}^\dagger =1.
\ee

We define new sets of canonical creation and annihilation operators by
\bea
c^\dagger = a^\dagger \frac{R+L}{2} + b^\dagger \frac{R-L}{2} ;&& d^\dagger = a^\dagger \frac{R-L}{2} + b^\dagger \frac{R+L}{2},\cr
c = \frac{R^\dagger+L^\dagger}{2} a + \frac{R^\dagger-L^\dagger}{2} b;&& d = \frac{R^\dagger-L^\dagger}{2} a + \frac{R^\dagger+L^\dagger}{2} b,\label{rplm}
\eea
and we can write
\be
{\cal H}_+ = -c^\dagger \Lambda c + d^\dagger \Lambda d.
\ee 

The ground states, $|0\pm\rangle$, are obtained by filling all the states corresponding to $c^\dagger$ and $a^\dagger$, respectively. Therefore, we have
\be
c^\dagger |0+\rangle =0;\qquad d|0+\rangle =0;\qquad \langle 0-| a=0;\qquad \langle 0-| b^\dagger =0.
\ee

Using Equation \eqref{rplm}, we can write
\be
a^\dagger = c^\dagger \frac{2}{R+L} - b^\dagger A;\qquad 
b = \frac{2}{R^\dagger+L^\dagger } d + G a,
\ee
where
\be
G = \frac{1-V}{1+V}.
\ee

Using the above equations, we can write
\be
\bar\chi b + a^\dagger \chi = Q_+ + Q_-,
\ee
where
\be
Q_+ = c^\dagger \frac{2}{R+L} \chi + \bar\chi \frac{2}{R^\dagger+L^\dagger} d;\qquad
Q_- = \bar\chi G a - b^\dagger G \chi.
\ee
Since
\be
Q_+ = a^\dagger \chi + b^\dagger G\chi -\bar\chi G a +\bar\chi b
\ee
it follows that
\be
[Q_+,Q_-] = -2\bar\chi G \chi.
\ee

Therefore, we have
\bea
Z(\eta,\bar\eta) &=& \int d\bar\xi d \xi e^{-\bar\xi \xi} \langle 0-| \exp [ Q_+ + Q_-] | 0 + \rangle 
\cr &=& \int d\bar\xi d \xi e^{-\bar\xi \xi} \exp\left ( \frac{1}{2} [ Q_+,Q_-]\right) \langle 0 - | e^{Q_-} e^{Q_+} |0+\rangle\cr
&=& \int d\bar\xi d \xi e^{-\bar\xi \xi} \exp\left ( \frac{1}{2} [ Q_+,Q_-]\right) \langle 0 - |0+\rangle\cr
&=& \int d\bar\xi d \xi \exp [ -\bar\xi\xi -(\bar \eta +m\bar \xi) G (\eta+\xi)] \det \frac{R+L}{2}\cr
&=& \exp \left[ -\bar\eta \frac{G}{1+mG}\eta\right] \det (1+mG) \det \frac{1+V}{2} \det R\cr
&=& \exp \left[ -\bar\eta \frac{1-V}{1+m+(1-m)V}\eta\right] \det \left[ \frac{(1+m)+(1-m)V}{2}\right] \det R.
\eea

The fermion mass is in the range $m\in [-1,1]$ and the fermion determinant is gauge invariant. There is a phase ambiguity present in the fermion determinant due to $\det R$
and the fermion determinant at $m=1$ is $\det R$. 
The choice of fixing this phase is tied to the 
choice of preserving parity symmetry at $m=0$ at the cost of 
introducing gauge anomaly, and the choice of preserving gauge invariance at the cost of losing parity symmetry at $m=0$.
For the latter option, 
the choice of $\det(R)=1$ fixes the phase of
infinite mass, $m=1$, fermion and preserves
gauge invariance for all values of $m$. In this paper, we will set $\det R=1$ from 
here on.

\section{Introducing Chern--Simons and Half-Chern--Simons Terms on the Lattice} \label{sec3}
With the choice of phase as explained in the last 
section, the fermion determinant becomes
\be
Z(m) = \det \left[ \frac{(1+m)+(1-m)V}{2}\right],
\ee
and it satisfies
\be
\frac{Z(m)}{Z^*(-m)} = \det V .
\ee

This is the parity anomaly which has the built in feature that if we write
\be
Z(m) = |Z(m)| e^{-i\Phi(m)}
\ee
then
\be
e^{-i[\Phi(m) + \Phi(-m)]} = e^{-2i\Phi(0)} = \det V,\label{csmassrel}
\ee
and
$\Phi(0)$ is usually written using the $\eta$-invariant as $\frac{\pi\eta}{2}$ (Note, the propagator satisfies 
$G(m)= -G^\dagger(-m)$ and preserve parity. Thus the anomaly 
is in the fermion induced gauge measure.)
With this lattice formalism, we have all the required ingredients 
for constructing Chern--Simons theories on lattice by the 
identification of the parity-odd phase of $\det(V)$ with 
level-1 Chern--Simons action.
As the simplest case, we can introduce a 
level-$k$ Chern--Simons action as 
\be
e^{i S_{\rm CS}(k)} = \det(V)^k.
\ee

In the massless limit, it is easy to see that the phase of 
$\det(1+V)$ is half of $\det(V)$ up to $\pm1$. We can introduce
the so-called $U_{k+N_f/2}(1)$ ``half-Chern--Simons'' theories 
on the lattice as 
\be
e^{iS_{\rm CS}(k+N_f/2)} = \det\left(\frac{1+V}{2}\right)^{N_f} \det(V)^k.
\ee

First, how do we know that $\det(V)$ is the same as Chern--Simons term?
In the study in \cite{Coste:1989wf}, it is analytically shown 
that the phase of $\det(X)$ in the massive Wilson fermion case 
is the same as Chern--Simons term. The pure phase $\det(V)$ 
in the case of overlap fermions is the same as the phase of $\det(X)$,
and, hence, we can borrow their results for overlap fermions.
In the subsequent two sections, we will also take an empirical 
approach and show that for cases of Abelian and non-Abelian 
background fields where Chern--Simons term can be exactly be worked out,
the phase $\det(V)$ indeed approaches the expectations in the 
continuum limit. Second, how did we manage to introduce 
``half-Chern--Simons'' term in an evidently gauge-invariant manner?
Using a non-trivial Abelian background in the next section, we
demonstrate this through the flow of the phase of massless overlap fermion determinant as a function of the Wilson loop, 
$e^{i 2\pi h_3}$ for $h_3\in[0,1]$, and show that at specific $h_3$ where there is a discontinuity in the phase at $m=0$, the determinant also
vanishes.

\section{Fermion Determinant in an Abelian Background with Uniform 
Magnetic Flux and Non-Trivial Temporal Wilson Loop} \label{sec4}

We now analyze the complex fermion determinant of a two-component three dimensional fermion in a well known Abelian background of interest both from the view point of showing subtle properties under gauge invariance and also from its relevance in condensed matter physics \cite{Seiberg:2016gmd}. The gauge field background on a continuum $\ell^3$ torus is
\be
A_1=-\frac{2\pi Q x_2}{\ell^2};\qquad A_2 = 0;\qquad A_3 = \frac{2\pi h_3}{\ell}.\label{fluxb}
\ee

Since $A_1(x_1,\ell,x_3)$ has to be gauge equivalent to $A_1(x_1,0,x_3)$, $Q$ has to be an integer. In addition, gauge invariance sets all $h_3+n$ to be equivalent for any integer $n$. 
The evaluation of the Chern--Simons action for this background in Equation \eqref{fluxb} is tricky \cite{Witten:2015aoa} and yields
\be
S_{\rm cs} = 2\pi h_3 Q.\label{scsflux}
\ee

Since $F = \frac{2\pi Q}{\ell^2} d x_1 \wedge dx_2$, for this background, $Q$ is the topological charge in all two-dimensional slices at a fixed $x_3$ and the deformation of $A$ to $tA$ has to connect two dimensional gauge fields in disconnected spaces. With a lattice regularization, $tQ$, as $t$ goes from $0 \to 1$, will result in $Q$ levels of the two dimensional Wilson--Dirac operator crossing zero \cite{Karthik:2015sza} and the phase within lattice regularization properly reproduces the gauge invariant Chern--Simons action \cite{Karthik:2015sza}. Overlap fermions can be used to study the complex fermion determinant strictly in the massless limit with the lattice regularization in place and we will show that the massless fermion determinant has a zero in the path connecting $h_3$ and $h_3+1$ for a fixed $Q$ enabling it to correctly reproduce
(1) a smooth function of $h_3$,
(2) that is gauge invariant under $h_3 \to h_3+1$,
(3) equal to half of $S_{\rm cs}$ in Equation \eqref{scsflux} at all values of $h_3$,
and (4) has a jump in the phase at the location of the zero of the fermion determinant.

We can implement the above Abelian background on the lattice 
by using the gauge-links as
\be
U_1(x)=\begin{cases} 1 & x_1\ne L-1 \cr e^{-i\frac{2\pi Q}{L} x_2} & x_1= L-1\end{cases};
\qquad U_2(x)=e^{i\frac{2\pi Q}{L^2} x_1};\qquad U_3(x) = e^{i\frac{2\pi h_3}{L}};
\ee
on a three dimensional periodic lattice defined by the points $x_1,x_2,x_3 \in [0,L-1]$ and
\be
U_\mu(x+L\hat\nu) = U_\mu(x);\qquad \mu,\nu=1,2,3.
\ee

Only the plaquettes in the $(1,2)$ plane have a non-zero flux and they are given by
\be
U_{12}(x) = U_1(x) U_2(x+\hat 1) U^*_1(x+\hat 2) U^*_2(x) = \begin{cases} \exp\left[i \frac{2\pi Q}{L^2}\right] & x \ne (L-1,L-1,x_3) \cr
 \exp\left[ i\frac{2\pi Q}{L^2}-i 2\pi Q\right] & x \ne (L-1,L-1,x_3) \end{cases}.
\ee
We note that the flux is not uniform and singular in the continuum limit if $Q$ is not an integer. Therefore, we will set $Q$ to be integers.

Since the gauge field background does not depend on $x_3$, one can go to momentum space in this direction. We will assume fermions obey antiperiodic boundary conditions in this direction.
Setting these momenta to be $\left[ \frac{2\pi k_3}{L}-\frac{\pi}{L}\right] $, $k_3\in [0,L-1]$, the operators $B$ and $D$ reduce to
\bea
B(k_3) &=& \frac{1}{2} \sum_{\mu=1}^2 (2-T_\mu -T_\mu^\dagger) + 2\sin^2\frac{\pi\left(h_3-\frac{1}{2}+k_3\right)}{L} - m_w;\cr
D(k_3) &=& \frac{1}{2} \sum_{\mu=1}^2 \sigma_\mu (T_\mu -T_\mu^\dagger) + i\sigma_3 \sin\frac{2\pi\left(h_3-\frac{1}{2} +k_3\right)}{L}.
\eea
with the gauge fields in the (1-2) plane being $U_1(x)$ and $U_2(x)$.
Let us denote the fermion determinant by $Z_L(h_3,Q,m;m_w)$ on the $L^3$ periodic lattice in this background and note that
\be
Z^0_L(h_3,Q,m;m_w) = Z^0_L(h_3+1,Q,m;m_w).
\ee

\begin {figure} [H]
\includegraphics[scale=0.45]{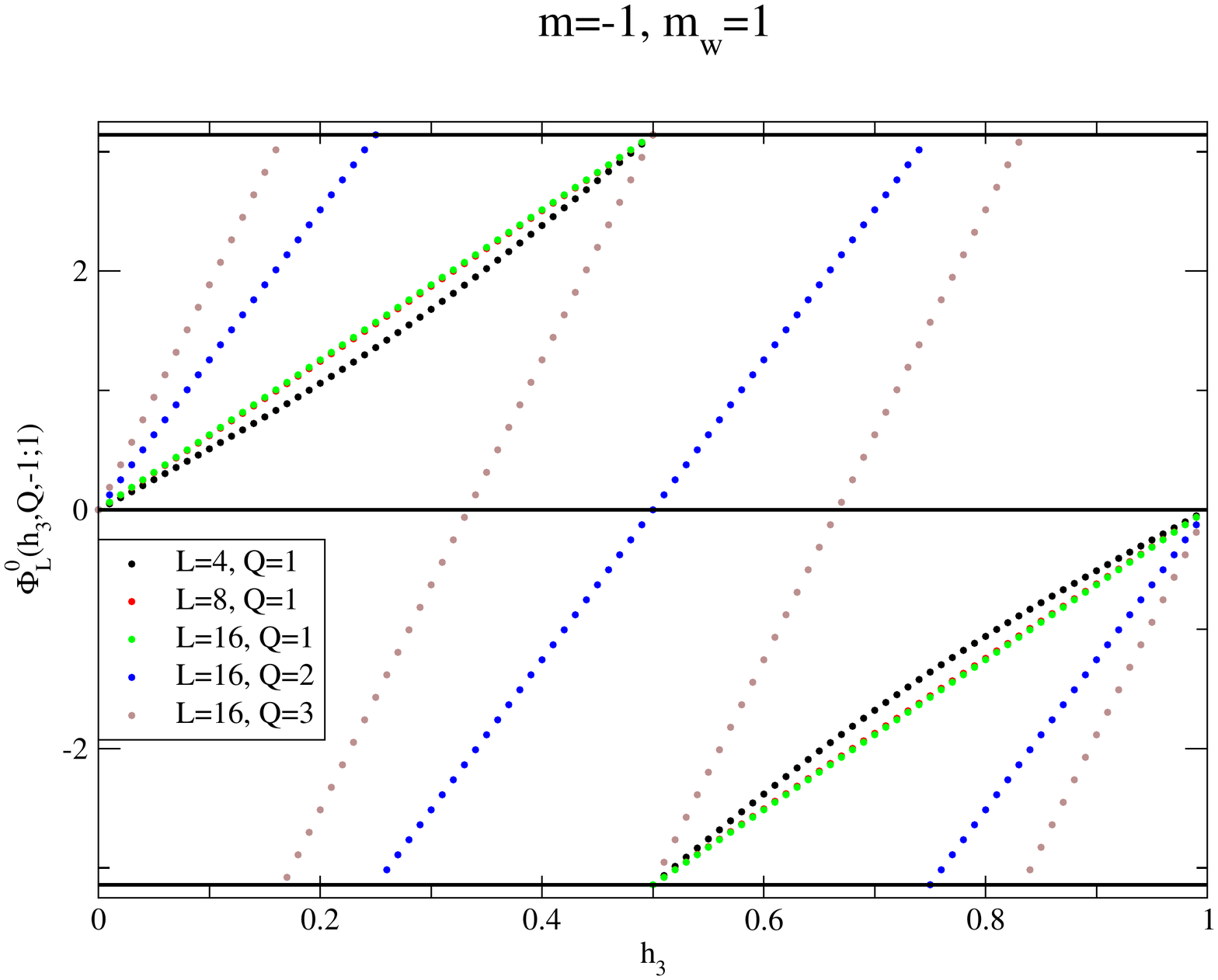}
\includegraphics[scale=0.45]{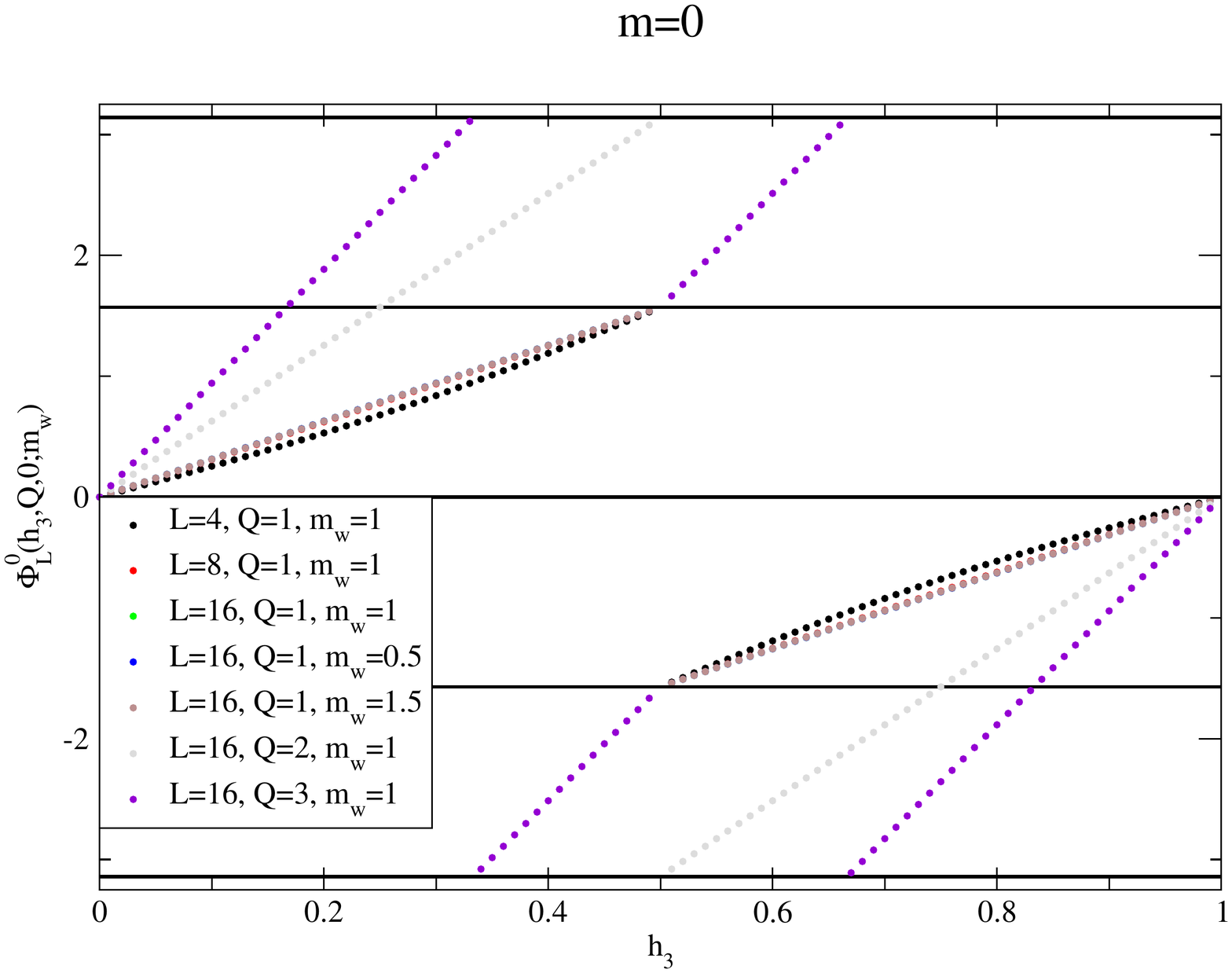}
\caption{The 
 top panel shows the flow of the phase 
$\Phi^0_L(Q,h_3)$ in the infinite mass case, $m=-1$, as a function of Wilson-loop variable 
$h_3\in[0,1]$ at $Q=1,2,3$. The Wilson mass entering the 
kernel of overlap operator is fixed at $m_w=1$. For fixed $Q=1$, the variation with 
reduction in lattice spacing by increasing $L$ from 4 to 8
is also shown.
The bottom panel shows similar flow of the phase of the
determinant in the massless case. The variability with respect to
the regulator parameter $m_w$ and lattice spacing are shown.
} \label{csu1}
\end {figure}

\begin {figure} [H]
\includegraphics[scale=0.5]{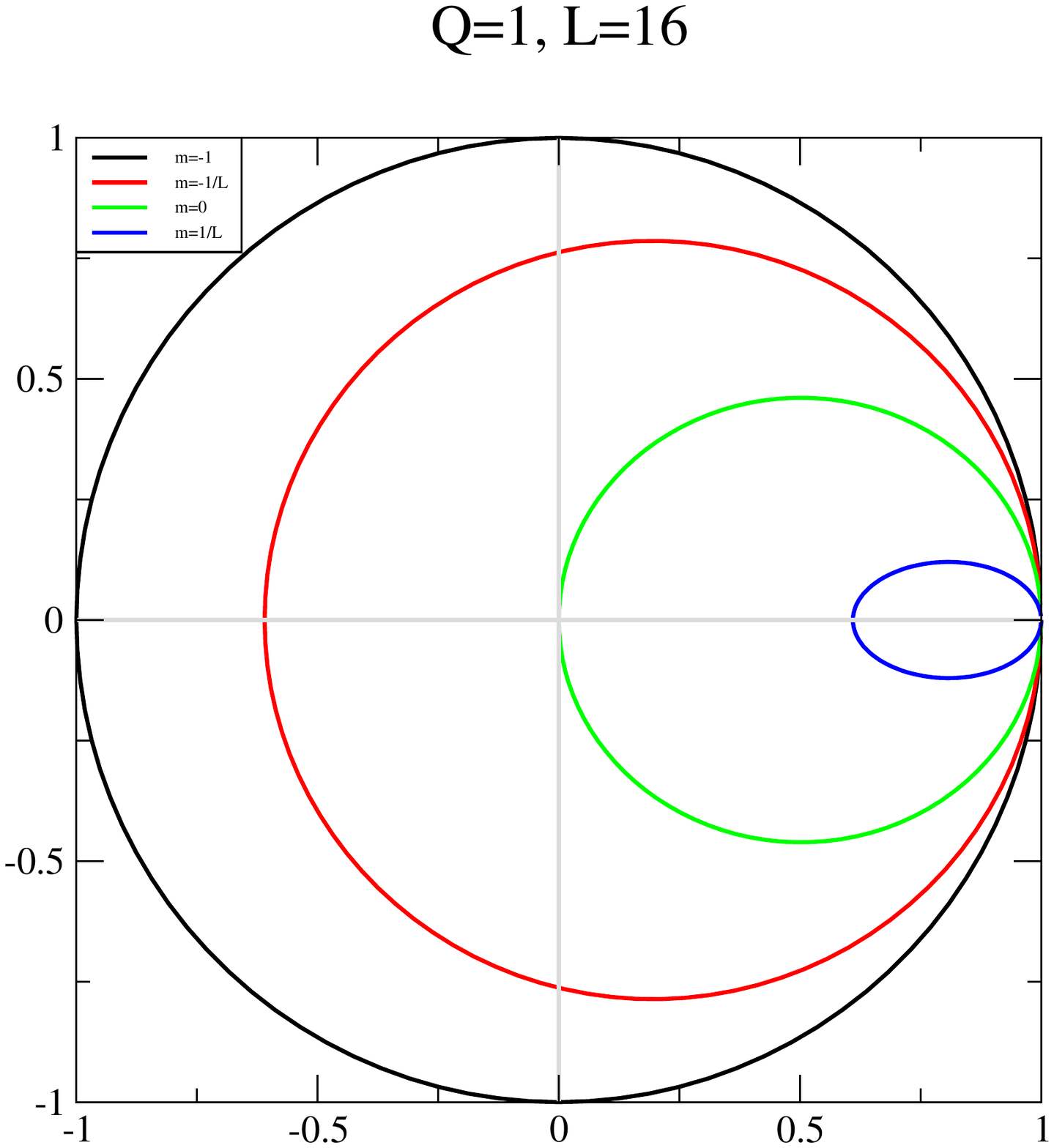}
\caption{The flow of the overlap fermion determinant in the 
complex plane as a function of $h_3$ at a fixed $Q=1$ on 
$L=16$ lattice. The flows are shown for $m= -$1 (black), $-1/L$ 
(red), 0 (green), $1/L$ (blue).
The flow starts at $(1,0)$ for $h_3=0$, goes clockwise
and returns back to $(1,0)$ for $h_3=1$.} \label{flowq1}
\end {figure}

\begin {figure} [H]
\includegraphics[scale=0.5]{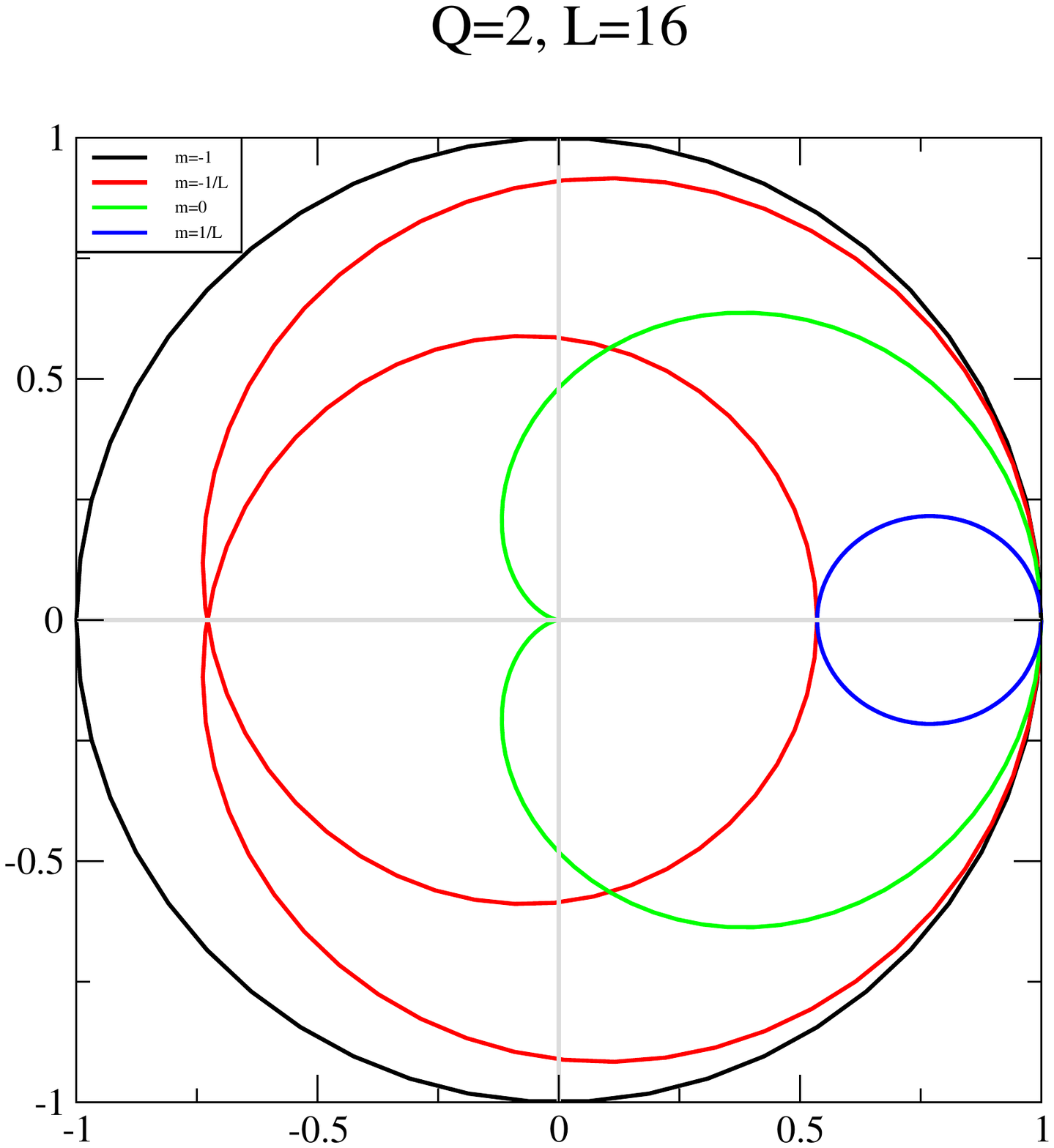}
\caption{The flow of the overlap fermion determinant in the 
complex plane as a function of $h_3$ at a fixed $Q=2$ on 
$L=16$ lattice. The description is the same as in 
Figure \ref{flowq1}.} \label{flowq2}
\end {figure}

\begin {figure} [H]
\includegraphics[scale=0.5]{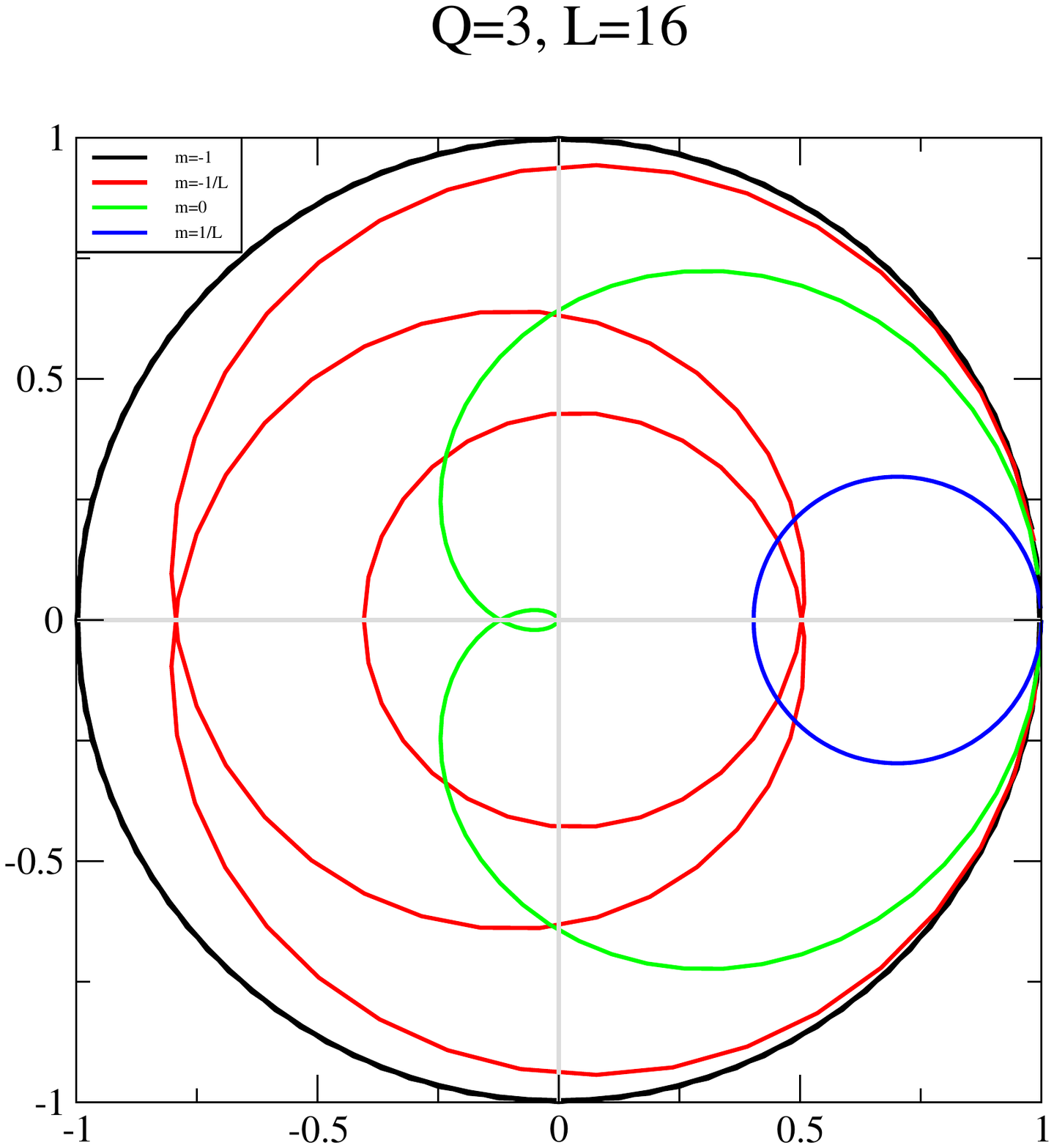}
\caption{The flow of the overlap fermion determinant in the 
complex plane as a function of $h_3$ at a fixed $Q=3$ on 
$L=16$ lattice. The description is the same as in 
Figure \ref{flowq1}.} \label{flowq3}
\end {figure}

\begin {figure} [H]

\includegraphics[scale=0.5]{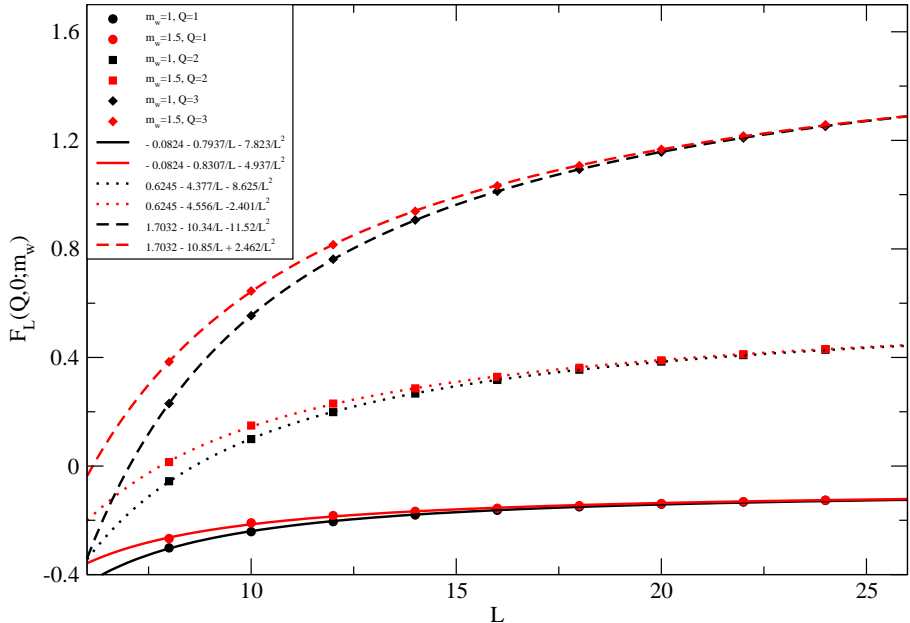}
\caption{
The plot demonstrates the existence of the continuum limit 
of the overlap fermion action in constant flux background at 
zero $h_3$. The continuum extrapolations ($L\to \infty$)
are shown using an expansion in lattice spacing $1/L$. 
The consistency in the extrapolated values using 
different regulator parameter $m_w$ is seen.
} \label{magu1}
\end {figure}

We define
\be
Z^0_L(h_3,Q,m;m_w) = \frac{Z_L(h_3,Q,m;m_w)}{Z_L(0,Q,m;m_w)} = \left|Z^0_L(h_3,Q,m;m_w)\right|\exp\left[ - i\Phi^0_L(h_3,Q,m;m_w)\right].
\ee
and
\be
 \frac{Z_L(0,Q,m;m_w)}{Z_L(0,0,m;m_w)} = \exp\left[-F_L(Q,m;m_w)\right]
\ee
as the determinant with reference to $h_3=0$ and the determinant at $h_3=0$ with respect to the free determinant, respectively and $F_L$ is a real function.
The key properties of the overlap fermion determinant are shown in the figures from Figures \ref{csu1}--\ref{magu1}.
Let us start with the top panel of Figure \ref{csu1} which focuses on the Chern--Simons action,
namely, $\Phi^0_L(h_3,Q,-1;m_w)$. We have shown the results only for $m_w=1$ but the $L\to\infty$ limit is independent of $m_w$ and we should find
\be
\lim_{L\to\infty} \Phi^0_L(h_3,Q,-1;m_w) = 2\pi h_3 Q.
\ee

The top panel clearly shows that the correct limit is approached for $Q=1$ as $L\to\infty$ ($L=8$ and $L=16$ fall on top of each other) and the dependence on $Q$ is also as expected and the overlap fermion correctly reproduces the first subtle properly and this is an obvious consequence of the same result with Wilson fermions seen in \cite{Karthik:2015sza}. We move on to behavior of the phase for the massless fermions in the bottom panel of Figure \ref{csu1}. We should find
\be
\lim_{L\to\infty} 2\Phi^0_L(h_3,Q,0;m_w) = \lim_{L\to\infty} \Phi^0_L(h_3,Q,-1;m_w) 
\ee
and
\be
\lim_{L\to\infty} \Phi^0_L(h_3,Q,0;m_w) =\lim_{L\to\infty} \Phi^0_L(h_3+1,Q,0;m_w).
\ee

This necessitates a jump in the phase when the flux quantum, $Q$, takes on odd values. First of all, we see that the phase has a limit when $L\to\infty$ as seen by comparing
the behavior for $L=8, m_w=1$ and $L=16, m_w=1$. Furthermore, the results for $L=16, m_w=0.5$ and $L=16, m_w=1.5$ are indistinguishable from $L=16, m_w=1$ showing the independence on the regulator parameter, $m_w$, as $L\to\infty$. Finally, we see that the phase shows a jump of $\pi$ at $h_3=\frac{1}{2}$ for $Q=1$ and $Q=3$. 

The plot of the full determinant, $Z_L^0(h_3,Q,m;1)$, is shown for $Q=1$, $Q=2$ and $Q=3$ in Figures \ref{flowq1}--\ref{flowq3}
respectively.
In these plots, $h_3 \in [0,1]$, and the motion along the closed curve is clockwise starting from the normalized value of $Z_L^0(0,Q,m;1)=1$. When $m=-1$, the closed curves are unit circles that wind $Q$ times and this is shown for reference in all three plots. We set $mL$ to be a constant when $m \in (0,1)$ to maintain a constant physical mass. On the one hand, we see that $Z_L^0(h_3,Q,m;1)$ winds around $Q$ times for $m < 0$ and its magnitude changes with $h_3$. On the other hand, we see that the phase of $Z_L^0(h_3,Q,m;1)$, reaches a maximum and minimum value in the range $\left(-\frac{\pi}{2},\frac{\pi}{2}\right)$ for $ m > 0$ and its magnitude changes with $h_3$. With the behavior in place for $m< 0$ and $m>0$, we see that $Z_L^0\left(\frac{1}{2},Q,0;1\right)$ is zero
and enables a jump in the phase for odd values of $Q$ with it being a smooth function of $h_3$. 
Finally, we show the results for $F_L(Q,m;m_w)$ in Figure \ref{magu1}. It remains finite as $L\to\infty$, which is the continuum limit of the 
background field, and independent of the regulator parameter, $m_w$.

\section{Fermion Determinant in a Non-Abelian \boldmath{$SU(2)$} Background
with Non-Zero \boldmath{$\Tr(A\wedge A \wedge A)$}} \label{sec5}

 The second background we will consider is a constant $su(2)$ background on a $\ell^3$ torus given by
\be
A_1 = \frac{2\pi q_1 t_1}{\ell};\qquad
A_2 = \frac{2\pi q_2 t_2}{\ell};\qquad
A_3 = \frac{2\pi q_3 t_3}{\ell};\qquad
\ee
where $t_\mu$ are the $su(2)$ generators in 
color space given by Pauli matrices, 
normalized such that $t_\mu t_\nu = \delta_{\mu\nu} + i \epsilon_{\mu\nu\rho} t_\rho$. In this case $q_i \in \left[0,\frac{\ell}{2}\right]$ are all gauge inequivalent and the Chern--Simons action reduces to
\be
S_{\rm cs} = 16\pi^2 q_1 q_2 q_3.
\label{nacs}
\ee

Contrary to the Abelian background the phase of the massless fermion determinant is simply given by $8\pi^2 q_1 q_2 q_3$ and we will show this to be the case.
Defining $Z_L(q_1,q_2,q_3,m;m_w)$ as the lattice regulated overlap fermion determinant on a $L^3$ periodic lattice with $m$ being the fermion mass and $m_w$ being another regulator parameter, we will show that both
$$
\lim_{L\to \infty} \frac{ Z_L(q_1,q_2,q_3,m;m_w)}{Z_L(q_2,q_2,0,m;m_w)};\qquad 
\lim_{L\to \infty} \frac{ Z_L(q_1,q_2,0,m;m_w)}{Z_L(0,0,0,m;m_w)} 
$$
are both finite and independent of the regulator $m_w$.
This constant SU(2) background can be introduced on the 
lattice as the link variables
\be
U_\mu = e^{i\frac{2\pi q_\mu t_\mu}{L}}.
\ee

We will consider this background on a three dimensional periodic lattice defined by the points $n_1,n_2,n_3 \in [0,L-1]$
and
\be
U_\mu({\bf n}+L\hat\nu) = U_\mu({\bf n});\qquad \mu,\nu=1,2,3.
\ee

All values of $q_\mu$ that remain finite as $L\to\infty$ are gauge inequivalent.

One can go to momentum space in all three directions and write
\bea
B^{ia,jb}({\bf k}) &=& b \delta^{ij}\delta^{ab} +
\sum_\mu \left[ s_\mu(q) s^a_\mu(k) \delta^{ij} t_\mu^{ab} \right];\cr 
D^{ia,jb}({\bf k}) &=& i \sum_\mu \left[ c_\mu(q) s^a_\mu(k) \sigma_\mu^{ij} \delta^{ab} \right] 
+i\sum_\mu \left[ s_\mu(q) c^a_\mu(k) \sigma_\mu^{ij} t_\mu^{ab} \right],
\eea
where
\bea
&& b=3-m_w - \sum_\mu \left[ c_\mu(q) c^a_\mu(k) \right];\cr
&& c_\mu (p) = \cos \frac{2\pi p_\mu}{L} \qquad s_\mu(p) = \sin \frac{2\pi p_\mu}{L};\cr
&& c^a_\mu (k) = \begin{cases} \cos \frac{2\pi k_\mu}{L} & \mu=1,2\cr
\cos \frac{2\pi k_\mu +\pi}{L} & \mu=3\end{cases};
\qquad s^a_\mu(k) = \begin{cases}
\sin \frac{2\pi k_\mu}{L} & \mu=1,2\cr
\sin \frac{2\pi k_\mu+\pi}{L} & \mu=3\end{cases}.
\eea

We have assumed anti-periodic boundary conditions for fermions in the $\mu=3$ direction.
The matrix $X(k)$ is given by
\be
X(k) = \begin{pmatrix} 
\alpha_1 & \alpha_3 & \alpha_4 & \alpha_5 \cr
\alpha_3^* & \alpha_2 & \alpha_6 & \alpha_4 \cr
-\alpha_4^* & \alpha_6 & \alpha_1^* & \alpha_3 \cr
\alpha_5 & -\alpha_4^* & \alpha_3^* & \alpha_2^* 
\end{pmatrix}
\ee
where
\bea
\alpha_1 &=& b+s_3(q) s^a_3(k)+ic_3(q) s^a_3(k) +is_3(q) c^a_3(k)\cr
\alpha_2 &=& b-s_3(q) s^a_3(k)+ic_3(q) s^a_3(k) -is_3(q) c^a_3(k)\cr
\alpha_3 &=& s_1(q) s^a_1(k) -i s_2(q) s^a_2(k)\cr
\alpha_4 &=& ic_1(q) s^a_1(k) +c_2(q) s^a_2(k)\cr
\alpha_5 &=& is_1(q) c^a_1(k) - is_2(q) c^a_2(k)\cr
\alpha_6 &=& is_1(q) c^a_1(k) + is_2(q) c^a_2(k).
\eea

\begin {figure} [H]

\includegraphics[scale=0.43]{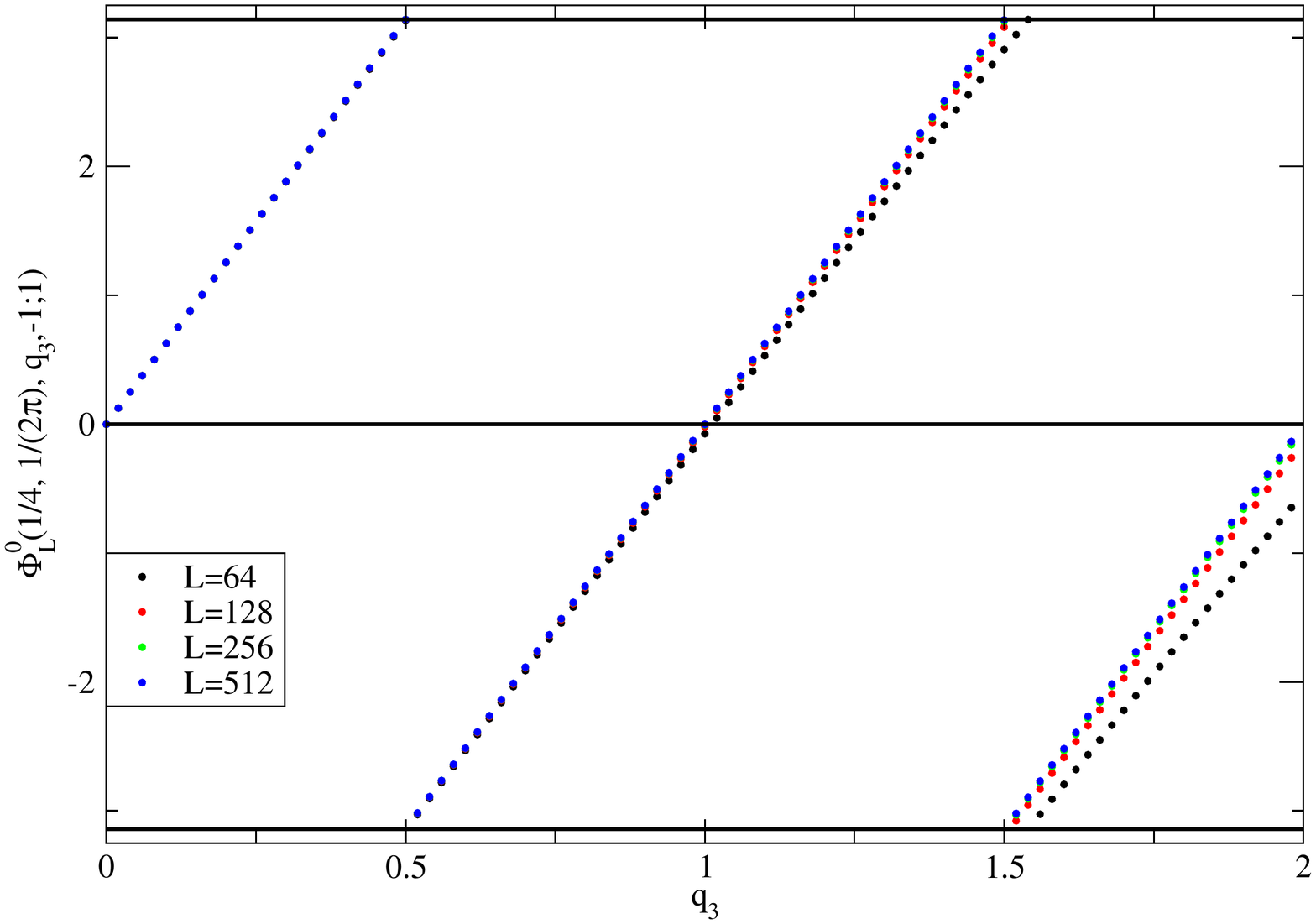}

\includegraphics[scale=0.43]{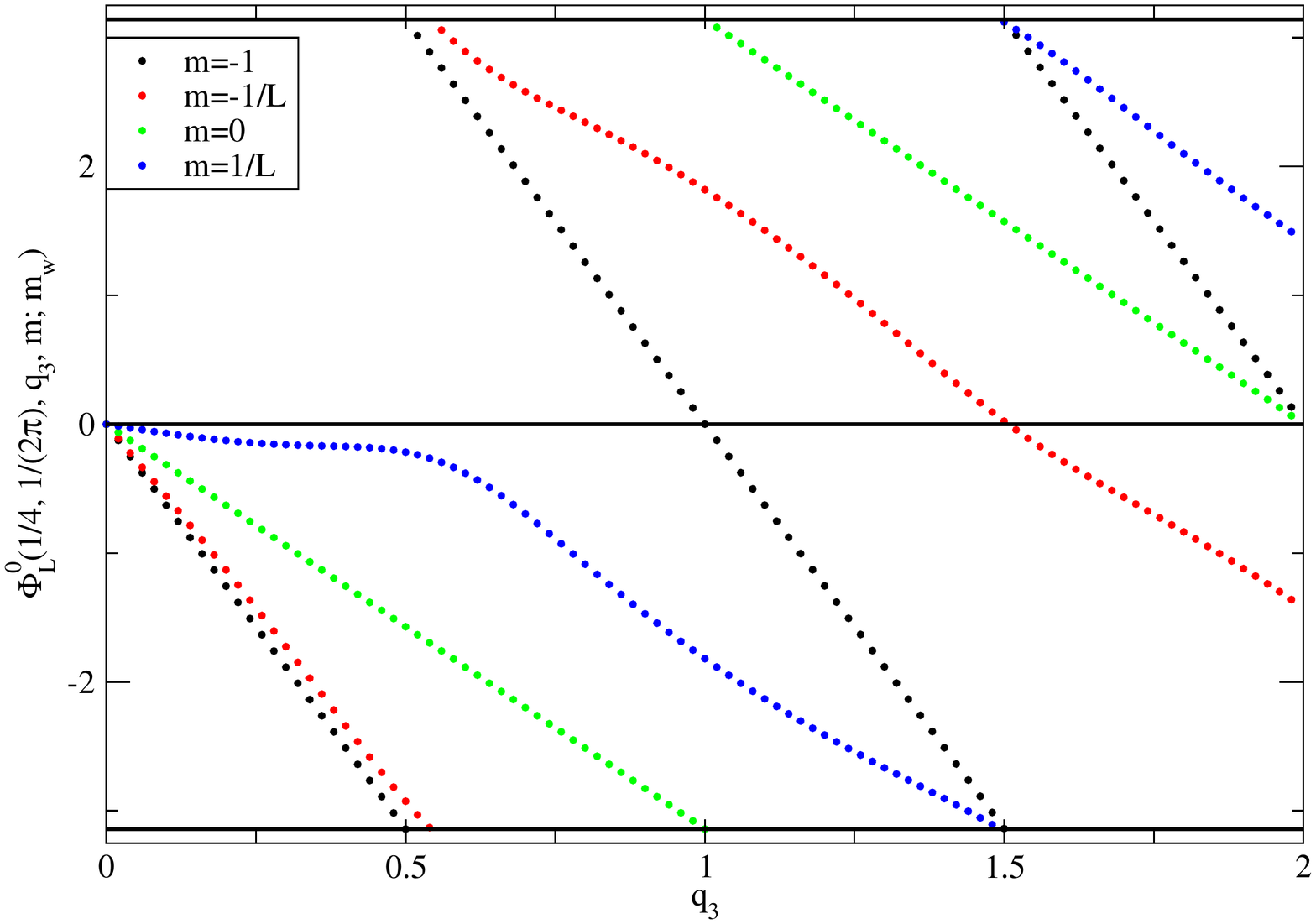}
\caption{
The figure is similar to Figure \ref{csu1} showing the flow of 
the phase of the fermion determinant as a function of 
SU(2) gauge field magnitude $q_3$. The top panel shows 
the result for infinitely massive fermion, $m=-1$ with 
regulator parameter $m_w=1$. The convergence of the results 
at different $L$ towards a continuum result is shown. 
The bottom panel shows the flow with $q_3$ at different 
fermion masses $m$.
} \label{su2flow}
\end {figure}

Let us denote the fermion determinant by $Z_L(q_3,m;m_w)$ on the $L^3$ periodic lattice in this background and define
\bea
Z^0_L(q_1,q_2,q_3,m;m_w) &=& \frac{Z_L(q_1,q_2,q_3,m;m_w)}{Z_L(q_1,q_2,0,m;m_w)} \cr
&=& \left|Z^0_L(q_1,q_2,q_3,m;m_w)\right|\exp\left[ - i\Phi^0_L(q_1,q_2,q_3,m;m_w)\right].
\eea
and
\be
 \frac{Z_L(q_1,q_2,0,m;m_w)}{Z_L(0,0,0,m;m_w)} = \exp\left[-F_L(q_1,q_2,m,;m_w)\right]
\ee
as the determinant with reference to $q_3=0$ and the determinant at $q_3=0$ with respect to the free determinant, respectively and $F_L$ is a real function.
We will set $q_1=\frac{1}{4}$ and $q_2=\frac{1}{2\pi}$ and vary $q_3$. The nonabelian Chern--Simons action given in Equation \eqref{nacs} reduces to $S_{\rm cs}=2\pi q_3$
and we show the phase of the overlap fermion correctly reproduces this result as $L\to\infty$ in the top panel of Figure \ref{su2flow}. Since all $q_3$ are gauge inequivalent,
we \mbox{should find}
\be
\lim_{L\to\infty} \Phi_L^0\left(\frac{1}{4},\frac{1}{2\pi},q_3,0;m_w\right) = \pi q_3
\ee
and we should also find
\be
\lim_{L\to\infty} \left[ \Phi_L^0\left(\frac{1}{4},\frac{1}{2\pi},q_3,\frac{m}{L};m_w\right) + \Phi_L^0\left(\frac{1}{4},\frac{1}{2\pi},q_3,-\frac{m}{L};m_w\right)\right]= 2\pi q_3
\ee

\begin {figure} [H]

\includegraphics[scale=0.45]{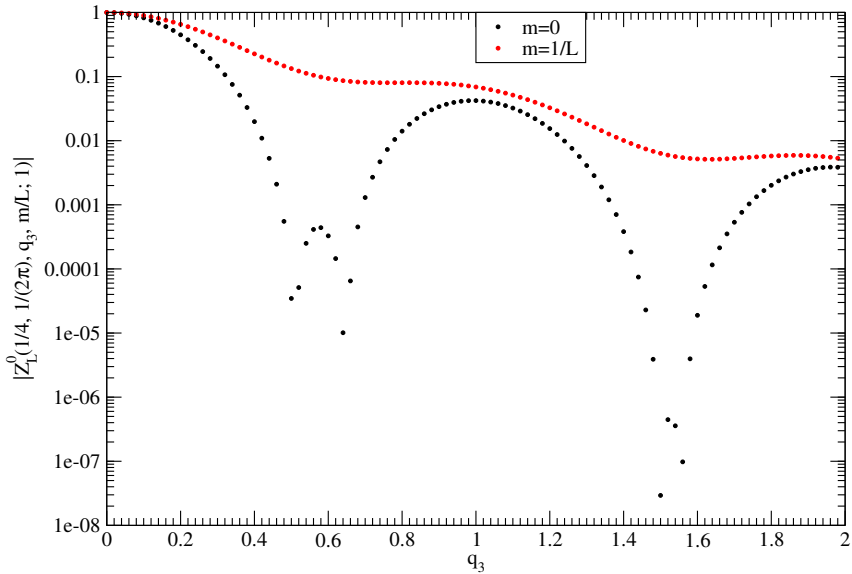}
\caption{
The 
dependence of the magnitude of the fermion 
determinant on $q_3$. The result at zero and 
non-zero masses are shown. The determinant at zero mass
vanishes are certain values of $q_3$, though not 
for any reasoning from invariance under large gauge transformation 
as seen in the case of Abelian background field studied in this 
paper.
} \label{su2zeros}
\end {figure}

Both these features are correctly reproduced in the top panel of Figure \ref{su2flow}. Since all $q_3$ are gauge inequivalent, we see that the phase at $q_3=1$ and $q_3=2$
only approaches $2\pi q_3$ as $L\to\infty$.
Note that unlike the Abelian case, the determinant winds around the origin for all values of fermion mass and the fermion determinant remains non-zero for all values of $q_3$. This is made clear through a plot of $\left| Z^0_L\left( \frac{1}{4},\frac{1}{2\pi},q_3,\frac{m}{L}; 1\right)\right|$
in Figure \ref{su2zeros}. 
\vspace*{-6pt}

We note a curious observation in this particular 
background. The fermion determinant for massless fermions becomes very small for certain values of $q_3$
and it has zeros even at finite $L$ that remains stable as $L\to\infty$ as seen in Figure \ref{su2zeros}. For our choice of $q_1$ and $q_2$, we find zeros a pair of zeros at
$q_3=0.508$ and $q_3=0.646$ and another pair at $q_3=1.502$ and $q_3=1.556$ that remain stable across $L$. 
In spite of the fact that all $q_3$ are gauge inequivalent, we see non-trivial behavior seen in the complex determinant for massless fermions in this particular background. 
Finally, similar to the Abelian background, we found the results for $F_L\left(\frac{1}{4},\frac{1}{2\pi},m;m_w\right)$ 
to be finite as $L\to\infty$ and independent of the regulator parameter, $m_w$.

\section{Mixed Chern--Simons (BF) Action and Dualities}\label{sec6}

Let $V_A$ denote the dependence of the unitary operator in Equation \eqref{voper} on the Abelian gauge field background, $A$. 
A {\sl mixed} Chern--Simons (BF) term can be written as
\be
 \det \left[ V_A V_B V^\dagger_{A+B} \right] = \det \left [ V^\dagger_A V^\dagger_B V_{A-B} \right] \sim e^{\frac{i}{2\pi} \int d^3 x \epsilon_{\mu\nu\lambda} A_\mu \partial_\nu B_\lambda}.
\label{zb1}
\ee

One can formally verify the identity by inserting the naïve expressions for CS that are only valid for perturbative fields.
The path integrals are defined over all gauge fields and a suitable measure such as a standard Maxwell action for gauge fields is needed
to verify the integrals non-perturbatively. Therefore, the last step
is essentially a mnemonic and it suggests relations of the form
\bea
\int [dA] \ \det \left [ V_{A-B} V^\dagger_{A-C} \right] &=& \delta(B-C),\cr
\int [dA] \ \det \left [ V_{A-B} V_{A-C} V^\dagger_{A-D} V^\dagger_{A-E} \right] &=& V_B V_C V^\dagger_D V^\dagger_E \ \delta(B+C-D-E),
\label{csdel}
\eea
using a gauge action for $A$ that is implicit and allows one to take a continuum limit (as pointed out explicitly in \cite{Karch:2016sxi}).

Dualities among various three dimensional theories start with the conjecture \cite{Seiberg:2016gmd}
that a theory with one massless two component fermion coupled to a dynamical gauge field $A$ and a classical background $C$ defined by
\be
Z(C) = \int [dA] e^{S_g(A)} \det\frac{1+V_A}{2} \det \left[ V_A^\dagger V_{A-C} \right] = \int [dA] e^{S_g(A)} \det\frac{1+V^\dagger_A}{2} \det V_{A-C} 
 \label{2p10},
\ee
is parity even and dual to a theory at the Wilson--Fisher fixed point. An explicit computation shows that
\bea
&&Z^*(C) = \int [dA] e^{S_g(A)} \det\frac{1+V_A}{2} \det V^\dagger_{A-C} 
= \int [dA] e^{S_g(-A)} \det\frac{1+V_{-A}}{2} \det V^\dagger_{-A-C} 
\cr
&\Rightarrow& Z^*(C) = \int [dA] e^{S_g(A)} \det\frac{1+V_A}{2} \det \left[ \left(V^\dagger_A\right)^2 \left(V^\dagger_C\right)^2 V_{A-C} \right] .
\eea

If $Z^*(C) = Z(C)$, we arrive at a non-trivial relation
\be
\langle \det V_A^\dagger \rangle = \det V_C^2
\ee
where the expectation value is with respect to the measure in Equation \eqref{2p10} and the lattice regularization can be used to verify this relation. In fact, if we use
\be
V_{2A} = V^4_A
\ee
which has been verified in the continuum limit when a measure for the gauge field is included \cite{Karthik:2018tnh}, we see that if we assume that the dynamical fermion has a charge of $2$ units,
\be
Z_2(C) = \int [dA] e^{S_g(A)} \det\frac{1+V_{2A}}{2} \det \left(V_A^\dagger\right)^2 \det \left[ V^\dagger_A V^\dagger_C V_{A-C}\right] \label{charge2}
\ee
then $Z^*_2(C) = Z_2(C)$ is trivially satisfied.

Regularized versions of the various duality relations discussed in \cite{Seiberg:2016gmd} can be obtained by following the steps found there. We multiply both sides of Equation \eqref{2p10} by
$\det V^\dagger_{B+C} $, promote $C$ to the dynamical field with $B$ being a background field and arrive at
a regularized version of a fermion-boson duality,
\be
e^{S_g(B)} \det\frac{1+V^\dagger_B}{2} 
= \int [dC] Z(C) \det V^\dagger_{B+C},\label{2p5}
\ee
after using Equation \eqref{csdel}.
If we assume $Z(C)$ is real we arrive at a regularized version of a boson--boson duality
\be
\int [dC] Z(C)\det \left [ V_B V^\dagger_{B+C} \right]
= \int [dC] Z(C) \det V_{B+C}. \label{2p6}
\ee

We can multiply both sides of Equation \eqref{2p10} by
$\det V_{B+C}$,
promote $C$ to a dynamical field with $B$ being a background field and arrive at
 a regularized version of a fermion-fermion duality 
\be
\int [dA] [dC] e^{S_g(A)} \det \frac{1+ V_A}{2} \det \left[ V_A^\dagger V_{A-C} V_{B+C} \right] 
= e^{S_g(B)} \det \frac{1+V_B}{2}\label{2p14}
\ee
and we have used Equations \eqref{2p5} and \eqref{2p6}.

A regularized version of a duality involving a fermion with charge of $2$ units discussed in \cite{Cordova:2017kue} can be obtained
by setting $B=-2X$ in Equation \eqref{2p14}. 
In this case, we can multiply both sides by ${\det }^2 V^\dagger_X$ to make the right-hand side even under parity.
We also multiply both sides by $\det \left [ V_X V_Y V^\dagger_{X-Y} \right] $ to couple it to an external flux and promote $X$ to a dynamical field.
Then we have
\bea
Z(Y) &= & \int [dA] [dC] [dX] e^{S_g(A)} \det \frac{1+ V_A}{2} \det V_A^\dagger \det \left[ V_{A-C} V_{-2X+C} \right] {\det }^2 V^\dagger_X
\det \left [ V_X V_Y V^\dagger_{X-Y} \right]\cr
&=& \int [dX] e^{S_g(2X)} \det \frac{1+V_{2X}}{2} {\det }^2 V^\dagger_X
\det \left [ V_X V_Y V^\dagger_{X-Y} \right]
\eea

Defining a change of variable, $X=Z+C$, in the first integral, we obtain
\be
Z(Y) = \int [dA] e^{S_g(A)} \det \frac{1+ V_A}{2} \det V_A^\dagger \det V_Y \int [dC] [dZ] \det \left[ V_{A-C} V_{2Z+C} 
V^\dagger_{Z+C} V^\dagger_{Z+C-Y} \right].
\ee

The integral over $C$ can be performed using Equation \eqref{csdel} This forces $A=Y$ and we arrive at the regularized version of a fermion-fermion duality 
\be
e^{S_g(Y)} \det \frac{1+ V_Y}{2} \int [dZ] \det \left [ V_Z V^\dagger_Y V_{Z+Y} \right]
= \int [dX] e^{S_g(2X)} \det \frac{1+V_{2X}}{2} 
\det \left [ V^\dagger_X V_Y V^\dagger_{X-Y} \right]
\ee
that connects a fermion with $2$ units of charge to a fermion with $1$ unit of charge.

We should remark that for the sake of simplicity and 
to a first degree of approximation, we assumed that 
the massless fermion limits of the odd-flavored theories considered above occurs at the 
``bare" fermion mass $m=0$. Unlike the parity-invariant 
theories with SU$\left(N_f\right)$ flavor symmetry 
with $N_f$ being even, where the mass term 
is protected by the symmetry, there is no such 
symmetry consideration in odd flavored theories.
Thus, it could be possible that one needs to tune 
the overlap fermion mass $m=m_c$ in order to reach 
criticality, provided there is one. In that case, the 
above set of equations might have to be modified accordingly with such mass terms, but it is a 
straightforward exercise.

\section{Conclusions}\label{sec7}
Overlap formalism was developed three decades ago \cite{Narayanan:1994gw} to properly reproduce all salient features of massless fermions in even dimensions. This was extended to odd dimensions in \cite{Kikukawa:1997qh} and we showcase the salient features of massless fermions in odd dimensions; particularly, we 
extended the formalism and spell out the lattice constructions of topological gauge actions that are being investigated 
currently in the context of TQFTs coupled to fermions, and 
in the context of infrared dualities. We focused on the overlap fermion determinant and used two examples, one Abelian background and one non-Abelian background. We showed that the overlap fermion determinant correctly reproduces all known properties of the phase of the fermion determinant; especially, we discussed how the lattice regularization manages
to implement the half-Chern--Simons term (or half-the-eta-invariant)
in a gauge-invariant manner.
While it is satisfying that we can nonperturbatively formulate 
the topological gauge theories on the lattice, an actual 
numerical study of such theories is not yet practical due to 
the sign problem and we did not address such issues in this 
paper.

An interesting possibility of having a lattice regularized Chern--Simons theory is the following. As we 
noted in this paper, it
is important to realize that the identification of overlap
$\det(V)$ with the continuum Chern--Simons 
action $e^{iS_{\rm CS}}$ is 
possible only in the continuum limit (as usual, the 
continuum limit taken at the 
trivial UV fixed point of the lattice gauge theory). 
However, as a lattice gauge theory that is away from 
any critical points, the overlap fermion determinants 
offer a great way to introduce new parity-odd gauge-invariant gauge-actions. Thus, one could now 
ask about the phase diagrams of such well defined 
lattice gauge 
theories as a function of different lattice couplings. 
This is an exciting direction to think about 
in the future.

\acknowledgments
R.N. acknowledges partial support by the NSF under grant number
PHY-1913010. N.K. is supported by Jefferson Science
Associates, LLC under U.S. DOE Contract \#DE-AC05-
06OR23177 and in part by U.S. DOE grant \#DE-FG02-
04ER41302. 
\bibliography{biblio}
\end{document}